\documentclass[manuscript]{emulateapj}

\usepackage{graphicx}

\usepackage{natbib}
\usepackage{url}
\usepackage{hyperref}
\hypersetup{colorlinks=false,pdfborder={0 0 0}}
\usepackage[utf8]{inputenc}
\usepackage[ngerman,english]{babel}
\newcommand{\hi}{H{\sc i}~}

\newcommand{\caii}{Ca{\sc ii}~}

\slugcomment{Submitted to The Astrophysical Journal Letters (ApJL)}

\shorttitle{The First Distance Constraint on the Renegade High Velocity Cloud Complex WD}
\shortauthors{}

\begin{document}


\title{The First Distance Constraint on the Renegade High Velocity Cloud Complex WD}


 \author{J. E. G. Peek\altaffilmark{1}, Rongmon Bordoloi\altaffilmark{2, 3}, Hugues Sana\altaffilmark{4}, Julia Roman-Duval\altaffilmark{5}, Jason Tumlinson\altaffilmark{5}, Yong Zheng\altaffilmark{6}}

\altaffiltext{1}{Space Telescope Science Institute, 3700 San Martin Dr., Baltimore, MD, 21218, USA. jegpeek@stsci.edu}
\altaffiltext{2}{MIT-Kavli Center for Astrophysics and Space Research, 77 Massachusetts Avenue, Cambridge, MA 02139, USA}
\altaffiltext{3}{Hubble Fellow}
\altaffiltext{4}{Institute of Astronomy, Celestijnenlaan 200d - box 2401, 3001 Leuven, NL}
\altaffiltext{5}{Space Telescope Science Institute, 3700 San Martin Dr., Baltimore, MD, 21218, USA.}
\altaffiltext{6}{Department of Astronomy, Columbia University, New York, NY, USA}

\begin{abstract}
We present medium-resolution, near-ultraviolet VLT/FLAMES observations of the star USNO-A0600-15865535. We adapt a standard method of stellar typing to our measurement of the shape of the Balmer $\epsilon$ absorption line to demonstrates that USNO-A0600-15865535 is a blue horizontal branch star, residing in the lower stellar halo at a distance of 4.4 kpc from the Sun. We measure the H \& K lines of singly-ionized calcium and find two isolated velocity components, one originating in the disk, and one associated with high-velocity cloud complex WD. This detection demonstrated that complex WD is closer than $\sim$4.4 kpc and is the first distance constraint on the +100 km s$^{-1}$ Galactic complex of clouds. We find that Complex WD is not in corotation with the Galactic disk as has been assumed for decades. We examine a number of scenarios, and find that the most likely is that Complex WD was ejected from the solar neighborhood and is only a few kpc from the Sun.%
\end{abstract}%



\section{Introduction}

High-velocity clouds (HVCs) provide a unique window into the coolest component of the circumgalactic medium and the processes of Galactic inflow and outflow. HVCs, and the complexes into which they are arranged, are found by their emission in \hi or absorption in numerous metal lines, and have radial velocities inconsistent with Galactic rotation \citep{Wakker_1997}. The precise origin of most HVCs is unknown, and some mix of Galactic fountain \citep[e.g.][]{Bregman_1980}, multiphase accretion \citep[e.g.][]{Fern_ndez_2012}, and gas stripping from satellites is typically invoked \citep{Putman_2012}. The exception is the Magellanic stream, which was stripped from the large and small Magellenic clouds, and which we will exclude from our discussion in this work. HVCs with negative radial velocities, which are metal enriched in the range of 10\% to 30\% of the solar metallicity, are likely a tracer of the process by which material accretes onto the Galaxy, though the total rate of this accretion is very uncertain. Less explored are the HVCs with positive radial velocities, most of which are in the inner two quadrants of the Galactic sky. These include the Wannier complexes WA, WB, WD, WE, and the Smith Cloud \citep{1991A&A...250..509W}. The Smith cloud has received significant attention of late, for its strongly cometary appearance which provides enough information to infer past trajectories, and make some inference as to its origin \citep{Lockman_2008, Fox_2015}.

Complex WD is the largest-area positive velocity HVC Complex, covering 310 square degrees with a total \hi flux of 1.2 $\times 10^7$ K km s$^{-1}$ arcmin$^2$, and a maximum \hi column density of $\sim 1.2 \times 10^{20}$ cm$^{-2}$. It is by far the largest complex that exists in the inner two Galactic quadrants, where a small fraction of HVC flux is detected. With a range of velocities between +90 and +130 km s$^{-1}$, it is consistent with cylindrical rotation on the far side of the inner Galaxy, 20 kpc from the sun with a mass of 6 $\times 10^7 M_\odot$. This would make it very similar in mass,  Galactocentric radius, and height to Complex C, the largest area and brightest HVC complex \citep{Thom_2008}. 

One major issue in gaining a better physical understanding of these enigmatic clouds is their unknown distance. Since there are no objects of fixed luminosity in HVCs, there are effectively no intrinsic distance measures. \hi emission or optical and UV absorption lines toward extragalactic background sources only provide distance-independent column densities. HVC distances not only give us a masses for these structures, but also a context; the spatial relationship between the cloud and the nearby spatial and kinematic structure of the disk gives us insight as to its origin.

There are a number of indirect methods for measuring the distance to an HVC complex, including H$\alpha$ emission and kinematic structure \citep{Putman_2003, Peek_2007}, but the only proven direct distance measure is stellar absorption. By observing stars with measured distances at medium or high spectral resolution, one can look for absorption lines in Na {\sc i}, \caii H \& K, Ti {\sc ii}, and numerous ultraviolet absorption lines at the velocity of \hi emission from HVCs \cite{1995A&A...302..364S}. By finding detections and non-detections of these absorption lines along lines of sight toward \hi emitting HVCs, distances can be robustly measured. A number of clouds have well-measured distances using this method, but complex WD is not among them \citep{Wakker_2001, Wakker_2007, Thom_2008, Wakker_2008}.

In this work we report the first distance upper limit on Complex WD using medium resolution absorption line spectroscopy toward a blue horizontal branch star. We extend the methods of \citet{Sirko_2004} to find the spectral type of the star, and thus put a precise distance limit. We use this to make some inferences as to the possible origin of Complex WD, and how it fits into the structure of Galactic HVCs as a whole.

\section{Data}

\subsection{New Observations}
The observations of our target, USNO-A0600-15865535, were obtained at the ESO Very Large Telescope (VLT) at Cerro Paranal, Chile on the nights of May 6th and May 7th, 2016. The target was observed as a part of our program ``Mapping the Cool Circumgalactic Medium with Calcium II'' (097.A-0552, PI: Peek) that uses the FLAMES/GIRAFFE spectrograph, a fiber-fed multiobject spectrograph mounted on the Nasmyth focus of UT2 \citep{Pasquini_2003}. USNO-A0600-15865535 is a bright ($g= 14.2$) very blue ($g-r = -0.21$) source, unresolved in Pan-STARRS imaging \citep{Magnier_2013, Schlafly_2012,Tonry_2012}. Through a combination of overoptimism and clerical error, it was targeted as a quasar candidate behind the circumgalactic medium of M83.


The GIRAFFE High Resolution mode was used in the H395.8 setup (HR02), which gives access to the 385.4 to 404.9nm wavelength range in the near UV at a spectral resolving power $\lambda / \delta \lambda$ of 22,700. A total of 4.5 h of integration were obtained, split in 3 exposures of 90 min. Given that most of the program targets were faint, the non-standard 50 khz,1$\times$1, high gain readout mode was used \citep{2008Msngr.133...17M}. Corresponding calibration frames were obtained  within 24h of the science observations. The raw scientific data were bias- and dark-subtracted, flat-fielded and wavelength-calibrated using the instrument's pipeline v12.14.2 under the esorex environment. The source spectra were extracted in the SUM mode. Finally the individual exposures were summed up to improve the signal-to-noise ratio of the extracted data.


\subsection{Archival Data}
\begin{figure*}[]
\begin{center}
\includegraphics[scale=3.5]{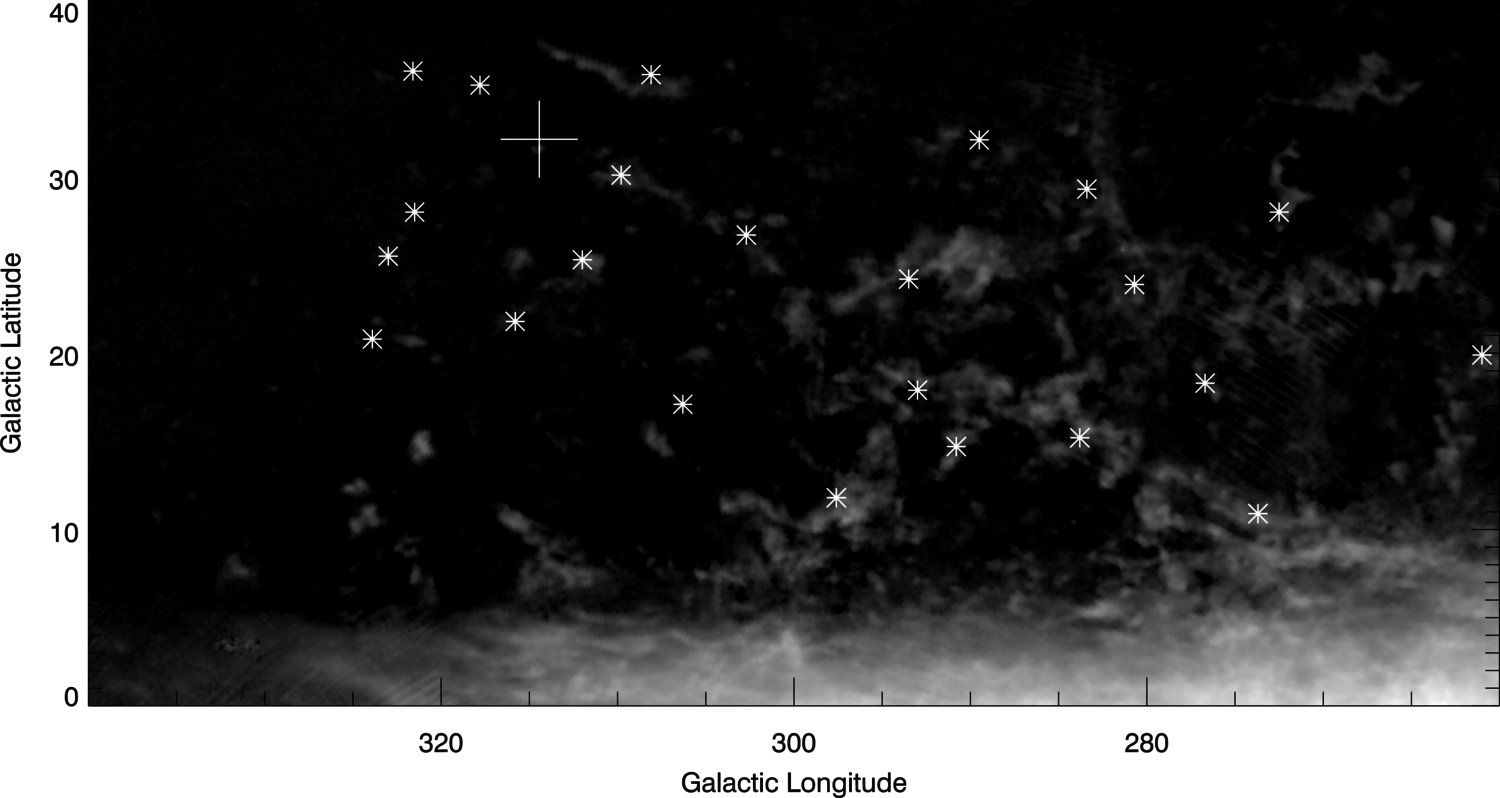}
\caption{{\label{fig:WD} Complex WD. The grayscale is \hi integrated column density between 98 km s$^{-1}$ and 131 km s$^{-1}$ on a log scale in Galactic coordinates. The feature below 5$^\circ$ is the Galactic plane, while the rest of the structure is Complex WD. The asterixes are the locations of the WD clouds in the Wakker van Woerden 1991 catalog. The cross is the location of USNO-A0600-15865535. The faint, parallel arc-like structures are systematic errors in the GASS data reduction.%
}}
\end{center}
\end{figure*}

In addition to the new data taken with FLAMES, we also used archival \hi data from the GASS survey \citep{Kalberla_2010}. GASS is a survey of Galactic \hi taken with the Parkes antenna, with a beam size of 16$^\prime$, a spectral resolution of 1 km s$^{-1}$, and an rms noise of 57 mK. These data allow us to map out the structure of Complex WD on the sky, and compare the spectrum of the \hi emission to \caii H \& K absorption along the line of sight. We present in Figure \ref{fig:WD} an image of Complex WD from GASS, along with the positions of clouds found in the \citet{1991A&A...250..509W} catalog, and the line-of-sight toward USNO-A0600-15865535.

\section{Methods \& Results}
\subsection{Stellar Classification}
The R $= 22700$, high SNR spectra of USNO-A0600-15865535 showed it to have very strong H$\epsilon$ and H$\zeta$ at $\sim 0$ km s$^{-1}$ in the local standard of rest (LSR) frame, and thus was very likely to be a halo star in the Milky Way, rather than the quasar we had originally attempted to target. Such stars are typically either blue horizontal branch (BHB) stars or blue stragglers (BS) stars, though can occasionally be hot main sequence (MS) stars.

\begin{figure}[h!]
\begin{center}
\includegraphics[scale=0.8]{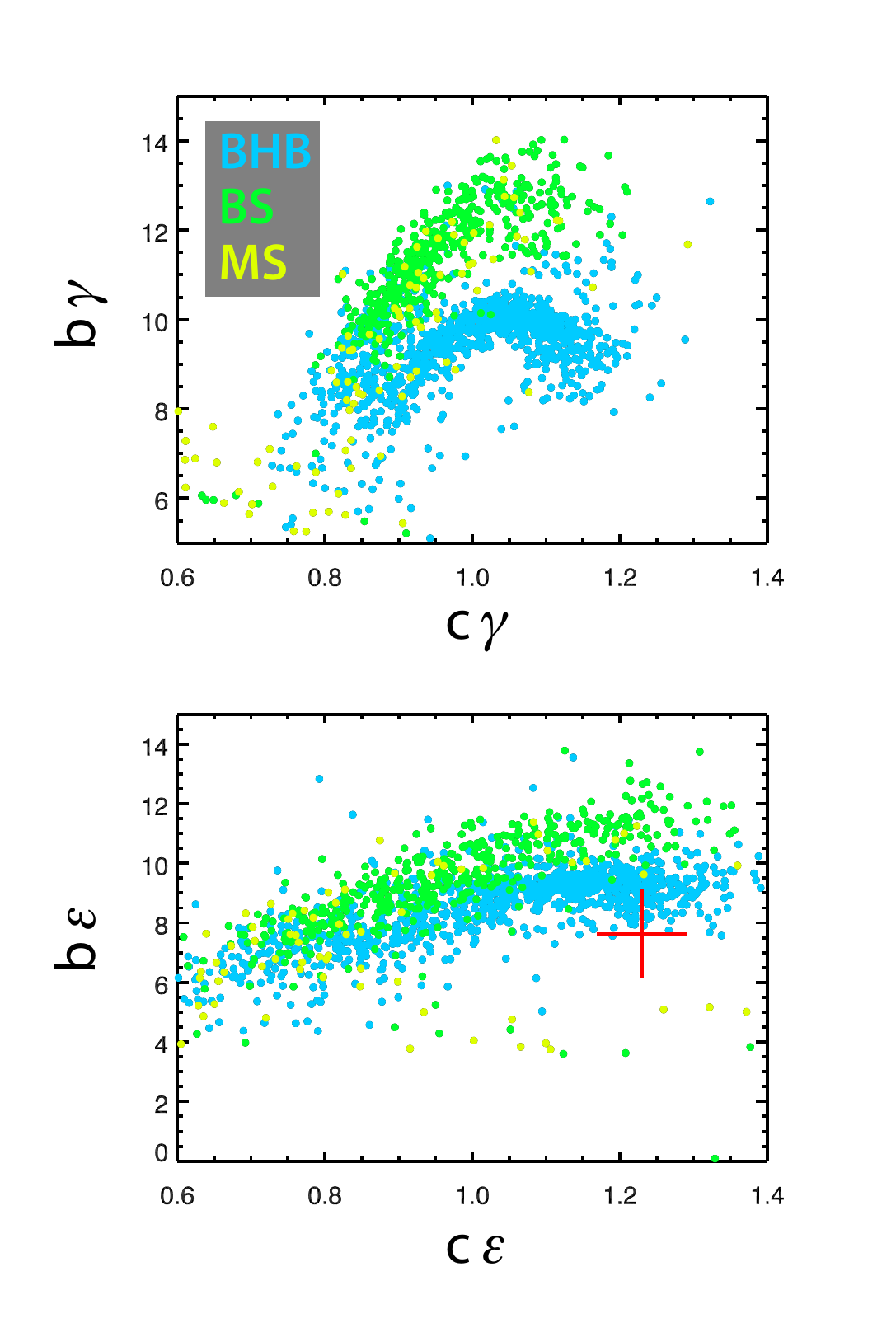}
\caption{{The Sersi\'c parameters of the H$\gamma$ (top) and H$\epsilon$ (bottom) lines as measured in 1637 stars from \citet{Xue_2008} with g $<$ 16. Overplotted in the bottom panel in a red cross is the result for the FLAMES spectrum of USNO-A0600-15865535, clearly a BHB star. \label{hgammahepsilon}%
}}
\end{center}
\end{figure}

To distinguish between these possibilities, we use a modification of the method developed by \citet{Sirko_2004} and later explored by \citet{Xue_2008}. In these works the shape of the H$\gamma$ and H$\delta$ lines are used to distinguish between these populations. These Balmer lines are fit with a standard Sersi\'c profile,

\begin{equation}\label{sersic}
\rm y = 1 - a~ exp \left[-\left(\frac{|\lambda-\lambda_0|}{b}\right)^{c}\right]
\end{equation}
and the parameters measured for these lines are distinct between the populations. Unfortunately, we do not have observations of the H$\gamma$ or H$\delta$ line, only the H$\epsilon$ and H$\zeta$ line. To determine whether we can use the (better resolved) H$\epsilon$ line as a similar discriminator, we first assemble a sub-population of BHB, BS, and MS drawn from the catalog presented in \citet{Xue_2008}, selected to be brighter than 16th magnitude in $g$. These targets have already been fit, and classified, but to test our method we refit the H$\gamma$ line in the SDSS DR10 spectra using Equation \ref{sersic}. We find that we can indeed reproduce the bifurcation between BHB and BS stars. We then apply this same fitting procedure to the H$\epsilon$ line. The results for both fits are shown in Figure \ref{hgammahepsilon}. The H$\epsilon$ line fit Sersi\'c parameters b and c do not as clearly delineate between BHB and BS stars, but the differentiation is still very much in place. We then apply this same fit to our VLT/FLAMES spectrum of USNO-A0600-15865535, and overplot the b and c parameters. USNO-A0600-15865535 is clearly a BHB star.

BHB stars have the useful property of being accurate standard candles. There is a direct relationship between BHB color and absolute magnitude reported in \citet{Xue_2008} (these values are originally reported in \citet{Sirko_2004} with a typographical error). The observed color in Pan-STARRS1 imaging is $g-r = -0.21$, and \citet{Schlegel_1998} report a reddening of $E\left(B-V\right) = 0.068$ toward this sightline, giving us a corrected color of $g-r = -0.28$ using the parameters from \citet{Schlafly_2011}. Using the correspondence reported in \citet{Sirko_2004}, this translates to an absolute magnitude of $M_g = 0.8$. USNO-A0600-15865535's measured $g$ magnitude of 14.2, reddening corrected to 14.0, gives a final distance of 4.4 kpc, with distance errors for BHB stars typically quoted at 10\%.

\subsection{ISM Line Measuement}

We perform Voigt profile fitting of the calcium II H \& K lines after dividing out the continuum spectrum and, in the case of the calcium II H lines, dividing out the Sersi\'c fit to the strong H$\epsilon$ line, which is centered at 230 km s$^{-1}$ LSR in the calcium II H frame. The column densities were determined by simultaneously fitting Voigt profiles to both lines of the \caii doublet with the \texttt{VPFIT} software \footnote{Available at \href{http://www.ast.cam.ac.uk/~rfc/vpfit.html}{http://www.ast.cam.ac.uk/$\sim$rfc/vpfit.html}}. For our Voigt profile fit analysis, the intrinsic model profiles are convolved with a Gaussian of FWHM 14 km s$^{-1}$ to account for the instrument resolution. The \caii line  close to the Milky Way rest frame is best fit with two Voigt profile components centered at  $v_{LSR} \,\approx$ -6 and 1 km s$^{-1}$, respectively. The redshifted high-velocity absorption component is best fit with two Voigt profile components centered at $v_{LSR}\, \approx$ 92.3 and 108 km s$^{-1}$, respectively. All four individual Voigt profile compoents to \caii K components agree with the equivalent \caii H parameters to within errors. The individual Voigt profile fits are reported in Table \ref{tab:vp} and are shown in Figure 3.

\begin{table}[ht]
\centering
\label{tab:vp}
\begin{tabular}{lcc}
b [km s$^{-1}$] & $v_{LSR}$ [km s$^{-1}$] & log(N [cm$^{-2}$]) \\
\hline
 6.3 $\pm$ 1.8 & -6.6 $\pm$ 0.6 & 12.05 $\pm$ 0.08  \\
20.7 $\pm$ 1.7 & 0.7 $\pm$ 1.7 & 12.15 $\pm$ 0.07  \\
 11.8 $\pm$ 22.2 & 92.3 $\pm$ 36.7 & 11.58 $\pm$ 0.80  \\
 10.4 $\pm$ 4.1 & 108.1 $\pm$ 7.1 & 12.17 $\pm$ 0.46  \\ 
\hline
\end{tabular}
\caption{{Voigt profile fit parameters for the Calcium II H \& K lines.}}

\end{table}

Three main \hi components are detected -- one near $v_{LSR} = 0$, one corresponding to the HVC at $v_{LSR} = 100$ and an intermediate velocity component. The \hi column density for each of these components are found by integrating the flux under the line under the optically thin assumption. We find a column density of $4.6 \times 10^{20}$ cm$^{-2}$, $2.2 \times 10^{19}$ cm$^{-2}$, and $8.6 \times 10^{20}$ cm$^{-2}$ for the low, intermediate, and high velocity components respectively.

\begin{figure}[h!]
\begin{center}
\includegraphics[scale=0.3]{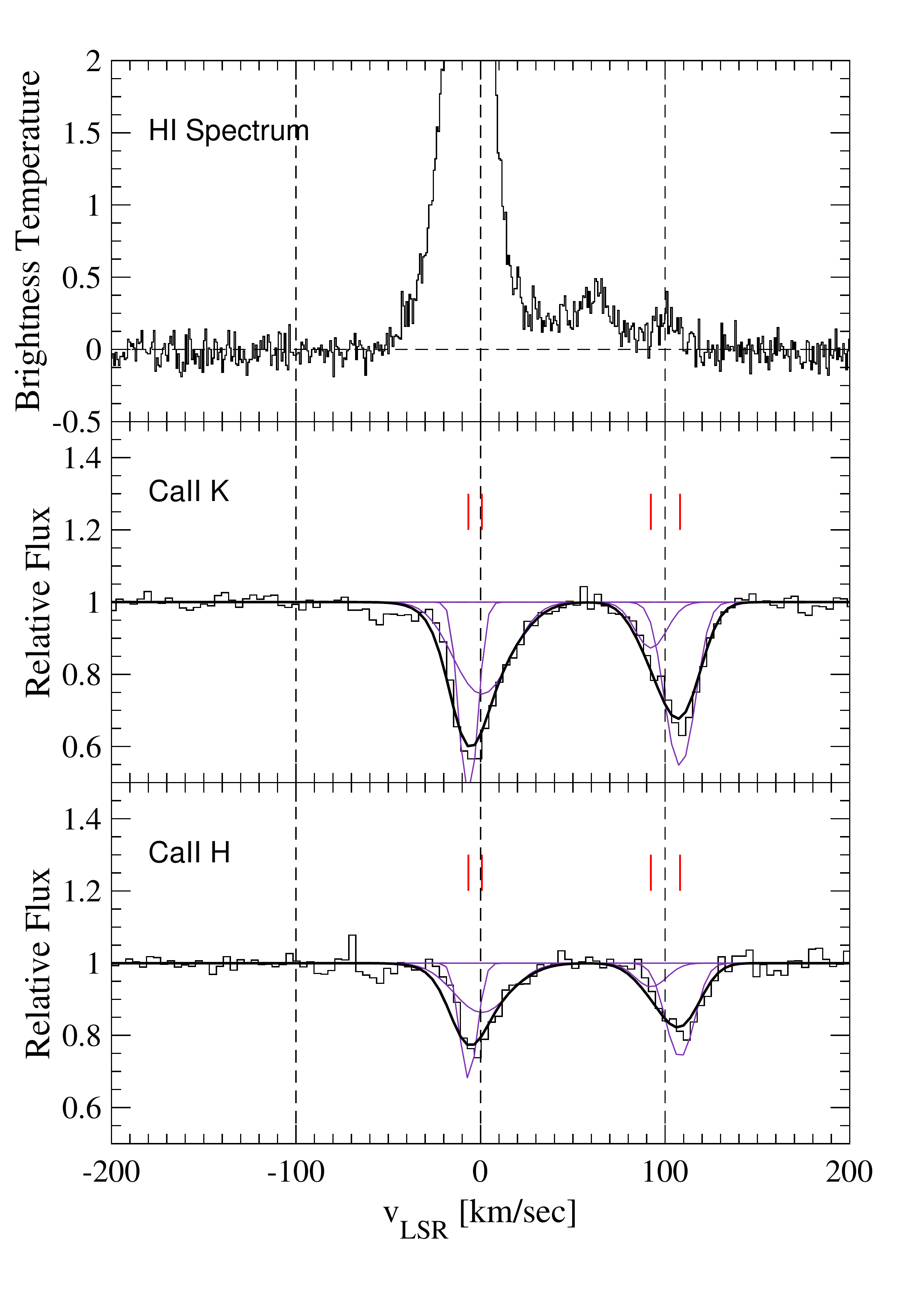}
\caption{{\hi and \caii toward USNO-A0600-15865535. The top panel is the GASS spectrum toward USNO-A0600-15865535, stretched to show the faint HVC feature at 100 km s$^{-1}$. The two bottom panels are the \caii K and H absorption line features, clearly showing the HVC absorption. The vertical red ticks show the location of the centroids of individual Voigt profile fits.%
}}
\end{center}
\end{figure}

\section{Discussion}

The clear detection of \caii H \& K absorption at +100 km s$^{-1}$ in USNO-A0600-15865535, coincident with \hi emission from Complex WD, indicates that the complex is closer than the star, at a distance 4.4 kpc. The column of \caii measured is comparable between the cloud and the disk, even though they have wildly different \hi columns. This is consistent with the weak \citep{Wakker_2000} or non-existent \citep{Bekhti_2012} correlation between \hi and \caii column density, which also explains the non-detection of the small intermediate velocity cloud along the line of sight at 60 km s$^{-1}$. Unfortunately, this lack of correlation makes it impossible to infer anything about the metallicity of the cloud from these metal absorption lines. Future metallicity measurements, perhaps toward USNO-A0600-15865535, will be critical in determining the origin of Complex WD, as gas of extragalactic origin typically has lower metallicities.

\begin{figure}[h!]
\begin{center}
\includegraphics[width=1.0\columnwidth]{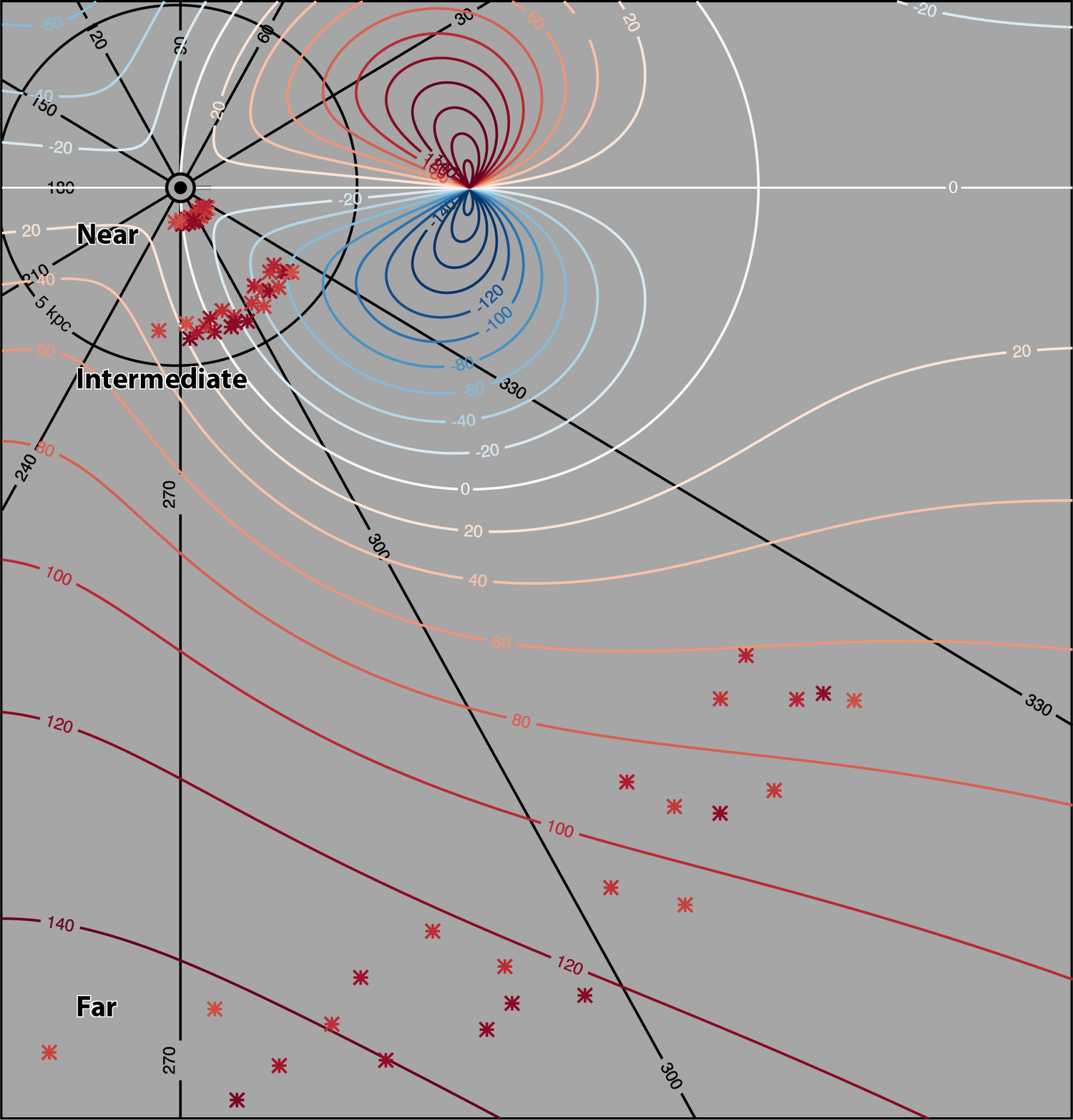}
\caption{{\label{fig:contour} A top-down view of the Galaxy. Lines of constant Galactic longitude are shown in black, meeting at the sun, marked by $\odot$. The colored lines represent the LSR velocity of the gas in the Galactic disk expected from a flat, 220 km s$^{-1}$ rotation curve, with bluer contours representing more negative velocities, and redder representing more positive. The asterisks are the Complex WD clouds found in \citet{1991A&A...250..509W}, colored according to their LSR velocities with the same scheme as the disk. We show three distance scenarios: the original ``far" scenario, which this work has ruled out, an ``intermediate" scenario, where all clouds are put at 4.4 kpc, and a ``near" scenario, where all clouds are 1 kpc from the sun.%
}}
\end{center}
\end{figure}

The kinematics and location of Complex WD do give us some clues as to its origin. Originally, \citet{1991A&A...250..509W} suggested that because complex WD is at positive velocity it is likely part of the structure of the Galaxy itself co-rotating with the disk. We call this the ``far" scenario in Figure \ref{fig:contour}, and it is emphatically ruled out by our detection of absorption. The appeal of the scenario is quite clear from the Figure -- a cloud at 20 kpc could be corotating with the disk. We now know that Complex WD is mostly inside the solar circle toward the fourth quadrant. Along the line of sight to USNO-A0600-15865535, Complex WD sits above a portion of the disk moving at $-$30 km s$^{-1}$ LSR if we assume that it is at the maximal distance of 4.4 kpc, decreasing to 0 km s$^{-1}$ LSR as we assume a closer distance. Complex WD is therefore strongly not in corotation with the disk. This is in rather stark contrast with other HVCs; a simplified model of HVCs with known distances found that they rotated with the disk at 77 km s$^{-1}$ -- slower than Galactic rotation, but with the same sense \citep{Putman_2012}. 

A number of scenarios could account for an overall difference in velocity between the cloud and the disk. An accreting cloud could easily have a much lower accretion velocity than the rotation speed of the disk, and the positive velocity observed could be an artifact of the solar motion. Similarly, it is possible that Complex WD is material ejected from star-forming regions closer to Galactic center \citep[e.g.][]{Ford_2010}, and thus the high positive velocity is an effect of the lower specific angular momemtum of that material. Both of these scenarios suffer from the fine tuning required to meet the very small LSR velocity gradient found in the complex. A flux-weighted first-order polynomial fit to the velocity gradient in the Wakker \& van Woerden 1991 catalog of WD clouds find $-0.072 \pm 0.146$ km s$^{-1}$ per degree of Galactic longitude. The reflex velocity of the solar motion represents 100 km s$^{-1}$ across 40 degrees of Complex WD -- unless the Cloud is conspiring to thwart our detection of a velocity gradient, we should see some effect of the solar motion. While we cannot fully rule out the ``intermediate" scenario, where the bulk of the cloud is at $\sim 4$ kpc, this velocity structure puts very tight constraints on any future model.

Finally we examine a ``near" scenario, where Complex WD is only 1-2 kpc away. In this scenario the cloud originated from an area near the sun, and thus has inherited the overall solar motion, largely solving the fine-tuning of the LSR velocity. In the ``near" scenario the cloud is ejected from the disk by some kind of impulsive event, perhaps connected to star formation in the Gould Belt or Saggitarus Arm, imparting an overall 100 km s$^{-1}$ bulk velocity. A cloud this far away would only be about 0.5 kpc above the disk, which is quite low for most known HVCs, and would make it distinct from all other known HVCs in its origin. The shearing effect of differential rotation is much weaker close to the sun, which makes this impulsive scenario more credible for a closer cloud. The somewhat symmetric Complex WE, with a similar velocity and location but at negative Galactic latitude, could conceivably have been generated by the same event.

\section{Conclusions}

In this work we have found that while the H$\epsilon$ line is not as powerful a discriminant between BHB, BS, and MS stars as the H$\gamma$ or H$\delta$ line, it can be used to show that USNO-A0600-15865535 is a BHB star approximately 4.4 kpc from the sun. We demonstrated that this star has clear Calcium H \& K absorption lines at a velocity coincident with Complex WD, and that therefore Complex WD must be closer than 4.4 kpc. We used this fact to rule out the originally assumed model of Complex WD, that is a Complex C-like cloud on the far side of the Galaxy, corotating with disk. Furthermore we investigated an intermediate distance scenario in which the Complex resides at $\sim 4$ kpc and found it difficult to reconcile the fixed LSR velocity of the cloud with the strong gradients implied from both the reflex solar motion and differential Galactic rotation. A "near" scenario, wherein a the complex was ejected from the solar vicinity to a distance of a few kpc seemed the most likely, though it would make the HVC the lowest in altitude known. 

Future observations toward USNO-A0600-15865535 would enable precise determinations of metallicities using other elements, which would help determine whether a disk-origin for this cloud is likely. Further observations towards closer stars, and across the face of the cloud could give us much more detailed information about the distance and three-dimensional morphology of the cloud, which would also help us understand how the cloud came to be, and how it relates to the accretion and feedback story of the Milky Way.

\section{Acknowledgements}

The authors thank Jessica Werk and Alis Deason for advice on the BHB spectroscopic determination method. This work is based  on observations made with ESO Telescopes at the La Silla Paranal Observatory under program ID 097.A-0552. This work used PanSTARRS1 data for targeting USNO-A0600-15865535 and accurate $g$ and $r$ band photometry. The Pan-STARRS1 Surveys have been made possible through contributions of the Institute for Astronomy, the University of Hawaii, the Pan-STARRS Project Office, the Max-Planck Society and its participating institutes, the Max Planck Institute for Astronomy, Heidelberg and the Max Planck Institute for Extraterrestrial Physics, Garching, The Johns Hopkins University, Durham University, the University of Edinburgh, Queen's University Belfast, the Harvard-Smithsonian Center for Astrophysics, the Las Cumbres Observatory Global Telescope Network Incorporated, the National Central University of Taiwan, the Space Telescope Science Institute, the National Aeronautics and Space Administration under Grant No. NNX08AR22G issued through the Planetary Science Division of the NASA Science Mission Directorate, the National Science Foundation under Grant No. AST-1238877, the University of Maryland, and Eotvos Lorand University (ELTE) and the Los Alamos National Laboratory. Partial support for this work was provided by NASA through Hubble Fellowship grant \#51354 awarded by the Space Telescope Science Institute, which is operated by the Association of Universities for Research in Astronomy, Inc., for NASA, under contract NAS 5-26555. 

\bibliographystyle{apj}

\end{document}